\documentclass{article}

\usepackage{makeidx}         
\usepackage{graphicx}        
\usepackage{multicol}        
\usepackage[bottom]{footmisc}
\usepackage[numbers]{natbib}

\newcommand{\bea}{\begin{eqnarray}}
\newcommand{\eea}{\end{eqnarray}}
\newcommand{\nn}{\nonumber}

\begin{document}

\begin{flushright}
KIAS-P18011
\end{flushright}

\vskip 1cm
\begin{center}
\centerline{\large \bf The cohomological structure of
generalized Killing spinor equations}
\vspace{1cm}

Dario Rosa \\

\bigskip
  School of Physics, Korea Institute for Advanced Study, Seoul 02455, Korea\\
\smallskip
Dario85@kias.re.kr

\abstract{We review the topological structure, sitting in any supergravity theory, which has been recently discovered in \cite{Imbimbo:2017}. We describe how such a structure allows for a cohomological reformulation of the generalized Killing spinor equations which characterize bosonic supergravity solutions with unbroken supersymmetry.}
\end{center}

\vskip .1in




\setcounter{footnote}{0}

\section{Introduction}
\label{sec:intro}

Localization has been a powerful tool to obtain exact results for supersymetric quantum field theories (SQFT) on curved spaces.\footnote{See \cite{Pestun:2016zxk} for an extensive overview.} To put a SQFT on a curved background preserving supersymmetry is a non-trivial task. 
A general strategy to address this problem\footnote{First considered in \cite{Buchbinder:1998qv}, using superspace formalism, and more recently re-discovered, using component formalism, starting from \cite{Festuccia:2011ws}.} is the following: one couples the SQFT under study to classical {\it off-shell} supergravity. Putting to zero the supersymmetry variations of the fermionic fields of supergravity one gets equations involving the bosonic supergravity fields. These equations, named { \it generalized  Killing spinor equations}, can be solved only for specific configurations of the supergravity background fields. We will refer to the space of these configurations as the {\it localization locus}.

In \cite{Imbimbo:2014pla} and \cite{Bae:2015eoa} the generalized Killing spinor equations for certain extended supergravity in two and three dimensions have been rewritten in a cohomological form. These cohomological equations were shown to be equivalent to the equations obtained setting to zero the BRST variations of the fermionic fields of topological gravity coupled to a given topological Yang-Mills system. A conceptual explanation of this equivalence has been furnished in \cite{Imbimbo:2017}. In this contribution we will review  this equivalence. The main technical tool we will use is the BRST formulation of supergravity, to which now we turn.

\section{The BRST formulation of supergravity}
\label{sec:BRSTsugra}

In the BRST formalism one introduces ghost fields, of ghost number $+1$, associated to each of the local symmetries. In supergravity, the {\it bosonic} local symmetries include diffeomorphisms and YM gauge symmetries; among the latter there are always local Lorentz transformations, plus additional local YM gauge symmetries which depend on the particular supergravity one is considering (a typical example is provided by the R-symmetry). We denote with $\xi^\mu$ the {\it anticommuting} vector ghost field associated to diffeomorphisms, and with $c$ the {\it anticommuting} scalar ghost field associated with the YM gauge symmetries, $c$ is valued in the adjoint representation of the total YM gauge algebra. The fermionic local symmetries are the $N$ local supersymmetries; for them one introduces {\it commuting} spinorial Majorana\footnote{We will refer to Majorana spinors for simplicity. The discussion can be extended, when $N$ is even, to Dirac spinors.} ghosts $\zeta^i$, with $i = 1 , \, \dots ,\, N$.

The spinorial ghosts $\zeta^i$, the vierbein $e^a \equiv e^a_\mu dx^\mu$ and the diffeomorphisms ghost $\xi^\mu$  constitute the {\it universal} sector of supergravity, in the sense that their BRST variations are the same for any supergravity theory 
\bea
\label{eq:universalSUGRA}
&& s\, \zeta^i = \iota_\gamma(\psi^i) + \mathrm{diffeos}+ \mathrm{gauge\; transfs} \ , \nn \\
&& s\, e^a = \sum_i \bar\psi^i \, \Gamma^a \, \zeta^i + \mathrm{diffeos}+ \mathrm{local\; Lorentz} \ , \nn \\
&& s \, \xi^\mu = - \frac 12 \, \mathcal{L}_\xi \xi^\mu -   \frac 12 \sum_i \bar \zeta^i \, \Gamma^a \zeta^i \, e_a^\mu =  - \frac 12 \, \mathcal{L}_\xi \xi^\mu + \gamma^\mu \ ,
\eea
where $s$ is the nilpotent BRST operator, $\psi^i \equiv \psi^i_\mu dx^\mu$ are the Majorana gravitinos, $\mathcal L_\xi$ denotes the Lie derivative along the vector $\xi^\mu$ and the vector $\gamma^\mu$ is the following bilinear\footnote{We will denote with $\gamma^\mu$ the vectorial bilinear (\ref{eq:gammadefinition}) and with $\Gamma^a$ the Dirac matrices.}
\bea
\label{eq:gammadefinition}
\gamma^\mu \equiv - \frac 12 \sum_i \bar \zeta^i \, \Gamma^a \zeta^i \, e_a^\mu \ ,
\eea
with $e^\mu_a$ the inverse of the vierbein. It was observed in \cite{Baulieu:1985md}, that the universal BRST variations (\ref{eq:universalSUGRA}) imply the following universal BRST variation for the vector bilinear $\gamma^\mu$
\bea
\label{eq:BRSTgamma}
s \, \gamma^\mu = - \mathcal L_\xi \, \gamma^\mu \ .
\eea

In \cite{Imbimbo:2017} it has been recognized that the universal BRST variations (\ref{eq:universalSUGRA}) and (\ref{eq:BRSTgamma}) precisely match the BRST variations of topological gravity, once one identifies the vector bilinear $\gamma^\mu$ with the superghost of topological gravity. Indeed, the BRST variations of topological gravity read
\bea
\label{eq:topgravBRST}
& s \,g_{\mu\nu} = - \mathcal L_\xi g_{\mu\nu} + \psi_{\mu\nu} \ ,  \qquad & s \, \psi_{\mu\nu} = - \mathcal L_\xi \psi_{\mu\nu} + \mathcal L_\gamma g_{\mu\nu} \ ,\nn \\
& s \,\xi^{\mu} = - \frac 12 \mathcal L_\xi \xi^\mu + \gamma^\mu \ , \qquad  & s \, \gamma^\mu = - \mathcal L_\xi \gamma^\mu \ , 
\eea
where $g_{\mu\nu}$ is the space-time metric, $\psi_{\mu\nu}$ is the topological gravitino and $\gamma^\mu$ is the topological gravity superghost. We have thus obtained that the universal sector of supergravity exactly coincides with topological gravity. We want now bring to light the full topological structure sitting inside any supergravity theory. 

\section{The full topological structure of  supergravity}

Beyond the ghost fields of ghost number $+1$ introduced in the previous section, any supergravity theory includes also fields of ghost number $0$. In the rest of this Section we will call both the fields of ghost number $0$ and the {\it commuting} supergravity ghosts $\zeta^i$ as the {\it matter} fields and we will denote them with $M$. 

The supergravity BRST variations of the matter fields read
\bea
\label{eq:BRSTvariationmatter}
s \, M = - \mathcal L_\xi \, M - \delta_c \, M + \hat M (M) \ ,
\eea
where $\delta_c$ is a gauge transformation with the ghost field $c$ and $\hat M (M)$ denotes a {\it composite} of the matter fields $M$ only. The expressions $\hat M (M)$, except for the universal supergravity fields discussed in Section \ref{sec:BRSTsugra}, are the {\it non-universal} parts of the supergravity BRST transformations; they are theory-dependent functionals of the matter fields. As an example, from (\ref{eq:universalSUGRA}) we find that for the universal fields $\zeta^i$, we have
\bea
\hat \zeta^i = \iota_\gamma \, \psi^i \ .
\eea

The BRST variations of the anticommuting ghost fields take a slightly different structure
\bea
\label{eq:BRSTvariationghost}
s \, \xi^\mu = - \frac 12 \, \mathcal L_\xi \, \xi^\mu + \gamma^\mu \ , \qquad s \, c = - c^2 - \mathcal L_\xi c + \hat c \ ,
\eea
where $\gamma^\mu \equiv \hat \xi^\mu$ is the vector bilinear (\ref{eq:gammadefinition}) and $\hat c$ are functions, of ghost number $2$, of the matter fields. The fields $\hat c$ are theory-dependent.

Imposing the nilpotency of the BRST operator $s$ on the matter fields $M$, one obtains the BRST rules for the composite $\hat M$ to be
\bea
\label{eq:sMhat}
s \, \hat M = - \mathcal L_\xi \hat M - \delta_c \hat M + \mathcal L_\gamma M + \delta_{\hat c} M \ .
\eea
The equations (\ref{eq:BRSTvariationmatter}) and (\ref{eq:sMhat}) make convenient to define another operator $S$, obtained by subtracting from $s$ both diffeomorphisms and YM transformations
\bea
S \, M \equiv s \, M + \mathcal L_\xi M + \delta_c M \ , \qquad S \, M = \hat M (M) \ . 
\eea
By applying $S$ on the composites $\hat M$ it follows
\bea
\label{eq:conditionsMhatone}
\frac{\partial \, \hat M}{\partial \, M} (M) \hat M (M) = S \, \hat M = S^2 \, M = \mathcal L_\gamma \, M + \delta_{\hat c} \, M \ ,
\eea
which defines a set of differential conditions that must be satisfied by $\hat M(M)$. Moreover, by computing $S^2 \hat M$ one gets
\bea\label{eq:S2Mhatone}
S^2 \hat M = \mathcal L_\gamma \, \hat M + \delta_{\hat c} \, \hat M   + \delta_{S \, \hat c} \, M \ ,
\eea
where the relation $S \, \gamma^\mu = 0$, which follows from (\ref{eq:BRSTgamma}), has been used.  On the other hand, since the fields $\hat M(M)$ are composite, and since the operator $S$ acts as a derivative, it must be
\bea\label{eq:S2Mhattwo}
S^2 \hat M = \mathcal L_\gamma \, \hat M + \delta_{\hat c} \, \hat M    \ .
\eea
By comparing (\ref{eq:S2Mhatone}) and (\ref{eq:S2Mhattwo}) one obtains that the composite $\hat c$ must satisfy the condition
\bea
S \, \hat c = 0 \ .
\eea

Hence, a  supergravity theory is specified by the composites $\hat M$ and $\hat c$, plus the universal composite $\gamma^\mu$ that has been discussed in the previous Section. On them one has to impose the constraints
\bea
\label{eq:constraintSUGRA}
&&  S \, \hat c = 0 \ , \nn \\
&& \frac{\partial \, \hat M}{\partial \, M} (M) \hat M (M) = \mathcal L_\gamma \, M + \delta_{\hat c} \, M \ .
\eea
When the constraints (\ref{eq:constraintSUGRA}) are imposed, the operator $S$ satisfies the algebra
\bea
S^2 = \mathcal L_\gamma + \delta_{\hat c} \ .
\eea

It can be shown (see \cite{Imbimbo:2017} for the details) that the composite $\hat c$ takes the general form
\bea
\label{eq:chatform}
\hat c = \iota_\gamma (A) + \phi \ ,
\eea
where $A$ is the gauge field associated to the local YM symmetry and $\phi$ is a scalar composite of the matter fields, bilinear in the supersymmetry ghosts $\zeta^i$ and valued in the adjoint of the YM Lie algebra. Its explicit form is {\it theory-dependent}.

The consistency condition $S \hat c = 0$ gets translated into the equation
\bea
S \, \phi = \iota_\gamma(S \, A) = \iota_\gamma (\hat A) \ .
\eea

The composite $S \, A = \hat A$ is the topological gaugino, usually denoted with $\lambda$. Together, the fields $\phi$ and $\lambda$ sit into a multiplet valued in the adjoint of the gauge algebra and whose BRST transformations are
\bea
\label{eq:topologicalYM}
&& S \, A = \lambda \ , \nn \\
&& S \, \lambda = \iota_\gamma \, (F) - D \, \phi \ , \nn \\
&& S \, \phi = \iota_\gamma \, (\lambda) \ ,
\eea
where $F$ is the field strength associated to the local YM symmetry.

The transformations (\ref{eq:topologicalYM}) are exactly the BRST variations of topological YM coupled to topological gravity, first derived in this form in \cite{Imbimbo:2009dy} and \cite{Imbimbo:2014pla}. This topological multiplet represents the universal topological sector sitting inside any supergravity theory.

Summarizing, the supergravity BRST algebra takes the universal form
\bea
\label{eq:BRSTalgebra}
S^2 = \mathcal L_\gamma + \delta_{\iota_\gamma (A) + \phi} \ ,
\eea
and it is characterized by the two topological fields $\gamma^\mu$ and $\phi$. The vector $\gamma^\mu$ has a universal form and it is identified with the superghost of topological gravity. The scalar $\phi$ has a theory-dependent form and it is identified with the superghost of topological YM coupled to topological gravity. We have thus identified the full topological content sitting inside any supergravity theory: the supergravity BRST algebra is characterized, universally, by two composite fields having clear topological roots.

\section{The cohomological equations of localization}
\label{eq:cohomological equations}

As mentioned in the Introduction, the localization locus of a given supergravity theory is obtained by setting to zero the supersymmetry variations of the fermionic supergravity fields. The resulting spinorial equations defining the localization locus are typically involved, and it is hard to extract their gauge invariant content. 

In the previous Sections it has been shown that a topological sector sits inside any supergravity theory. In particular, the composite topological fermions $\psi_{\mu\nu}$ and $\lambda_\mu$ have been constructed. Hence, on the localization locus the following equations must hold
\bea\label{eq:universalequations}
S \, \psi_{\mu\nu} = \mathcal L_\gamma g_{\mu\nu} = 0 \ , \qquad S \, \lambda = D \, \phi - \iota_\gamma (F) = 0 \ ,
\eea
since both $\psi_{\mu\nu}$ and $\lambda_\mu$ are composites containing the fermionic supergravity fields.
The first equation in (\ref{eq:universalequations}) states that the vector bilinear $\gamma^\mu$ has to be an isometry of the spacetime metric $g_{\mu\nu}$. This equation is indeed well-known in the supergravity literature.

On the other hand, the second equation is novel and it has not been studied extensively in both supergravity and topological field theory literature.\footnote{The author has been informed that this same equation is currently under investigation in a slightly different context \cite{Jacob:2017}.} This equation, when the YM gauge symmetry is non-abelian, is not gauge invariant: its gauge invariant content is captured by considering the following generalized Chern classes
\bea
c_n( F  + \phi) \equiv \mathrm{Tr}\,  (F  + \phi)^n \ . 
\eea 
Indeed, the generalized Chern classes $c_n$ satisfy the equations
\bea
\mathcal D_\gamma \, c_n \equiv (d - \iota_\gamma)\, c_n =0 \ ,
\eea
which states that the $c_n$'s, on the localization locus, are closed under the coboundary operator
\bea
\mathcal D_\gamma \equiv (d - \iota_\gamma) \ , \qquad \mathcal D_\gamma^2 = 0 \ , 
\eea
associated to the de Rham cohomology of forms on space-time, {\it equivariant} with respect to the action of the Killing vector $\gamma^\mu$. In the following, forms closed under the operator $\mathcal D_\gamma$ will be called {\it $\gamma$-equivariant}.

It should be stressed that the equations (\ref{eq:universalequations}) are {\it universal}, in the sense that they have to be satisfied, with a specific $\phi$ which is theory-dependent, on the localization locus of any supergravity theory. 

It should be also stressed that the equations (\ref{eq:universalequations}) in general do {\it not} completely specify the localization locus. Indeed they are obtained by setting to zero the supergravity BRST variations of specific (fermionic) supergravity bilinears, and there might be inequivalent bosonic supergravity backgrounds that give rise to $c_n$'s which are different representatives of the same $\gamma$-equivariant classes. As a matter of facts, the $\gamma$-equivariant classes $c_n$ parametrize different {\it branches} of the localization locus. On each of these branches, a moduli space of inequivalent solutions of the generalized Killing spinor equations can be usually found. 

In the following, other independent and gauge invariant composite fermions, which can be defined for specific supergravities only, will be introduced. Setting to zero their BRST variations one obtains additional cohomological equations which must be satisfied on the localization locus. These equations allow for a finer classification of the localization locus, i.e. they allow to characterize the moduli space sitting inside each of the branches defined by the $c_n$'s. 

To see how to extract these additional equations, one observes that the crucial property of $\phi$, which made possible to construct the topological multiplet $F + \lambda + \phi$ satisfying on the localization locus the second equation in (\ref{eq:universalequations}), is that its BRST variation is
\bea
S \, \phi = \iota_\gamma (\lambda) \ .
\eea

We note that also the supersymmetry ghosts $\zeta^i$ have a BRST variation of the same kind:
\bea
S \, \zeta^i = \iota_\gamma (\psi^i) \ .
\eea
Hence, scalar and gauge invariant ghost bilinears which are {\it independent} of extra bosonic fields automatically give rise to other topological multiplets whose BRST take the form (\ref{eq:topologicalYM}) and so, putting to zero the BRST variations of the corresponding fermions, one gets additional cohomological equations which are satisfied on the localization locus.

To provide an example, we wil consider the case of $N = (2,2)$ supergravity in two dimensions.  In $N= (2,2)$ $2$d supergravity, it  is convenient to combine the two Majorana spinors $\zeta^i$, $i = 1 , \, 2$ into a single Dirac spinor $\zeta$, on which the R-symmetry gauge group $U(1)$ acts as a phase multiplication. One can then construct the two scalar bilinears\footnote{Barred spinors are defined in the usual way: $\bar \zeta \equiv \zeta^\dagger \, \Gamma_0$.}
\bea
\label{eq:phi12}
\varphi_1 \equiv \bar \zeta \zeta \ , \qquad \varphi_2 \equiv \bar \zeta \, \Gamma_3 \, \zeta , 
\eea 
which are gauge invariant. Therefore, their BRST variations read
\bea
S \, \phi_i = \iota (\lambda_i) \ , \qquad i = 1, \, 2 \ ,
\eea
where
\bea
\lambda_1 \equiv \bar\psi \zeta + \bar \zeta \psi \ , \qquad \lambda_2 \equiv \bar\psi \, \gamma_3 \, \zeta + \bar \zeta \Gamma_3 \psi \ .
\eea
As consequence,  the BRST algebra (\ref{eq:BRSTalgebra}) tells that  the generalized forms
\bea
\mathcal{H}_i\equiv \phi_i+ \lambda_i + \hat{H}^{(2)}_i \ ,
\eea
satisfy
\bea\label{eq:descent}
 (S + d -\iota_\gamma) \, \mathcal{H}_i=0 \ .
\eea
The 2-forms $\hat{H}^{(2)}_i$ write
\bea
\hat{H}^{(2)}_1 = \bar\psi \, \psi + H^{(2)}_1\qquad \hat{H}^{(2)}_2 =\bar\psi \, \Gamma_3\,\psi + H^{(2)}_2 \ ,
\eea
where $H^{(2)}_i$, with $i=1,2$, are the graphiphoton field strengths. Note that these $2$-forms are non universal: they depend indeed on the auxiliary fields of $N=(2,2)$ supergravity and their explicit form can be found in \cite{Imbimbo:2017}. 
From (\ref{eq:descent}) one deduces that  on the localization locus the following cohomological equations hold
\bea
d\, \varphi_i - i_\gamma(H^{(2)}_i)=0 \ .
\label{ghostbieqs}
\eea

It has been shown\footnote{In Euclidean signature and with the space-time topology of the sphere $S^2$.} in \cite{Bae:2015eoa} that the equations (\ref{ghostbieqs}), together with the universal equations (\ref{eq:universalequations}), fully characterize the localization locus of $N = (2,2)$ supergravity: the localization locus splits in three branches which are parametrized by the integer values of the flux of the R-symmetry field strength; on each branch the equations (\ref{ghostbieqs}) give rise to a moduli space of inequivalent supersymmetric supergravity backgrounds. This moduli space is parametrized by two real moduli.

\section*{acknowledgement}
I am particularly grateful to C.~Imbimbo for a long collaboration on this subject over the years. I also thanks J.~Bae and J.~Winding for discussions and collaboration. I finally thanks V.~Dobrev and all the organizers of the ``X. International Symposium
QUANTUM THEORY AND SYMMETRIES'' for the invitation to this very interesting workshop.


\end{document}